\documentclass[]{iopart}

\usepackage{graphicx}
\usepackage{amsfonts}
\usepackage{iopams}
\usepackage{stackrel}
\usepackage{color}

\begin{document}

\def\tr{\mathop{\mathrm{Tr}\,}\nolimits}

\title{Glass phenomenology in the hard matrix model}

\author{
  Junkai Dong, Veit Elser, Gaurav Gyawali, Kai~Yen Jee, Jaron Kent-Dobias,
  Avinash Mandaiya, Megan Renz, and Yubo Su
}

\address{
  Laboratory of Atomic \& Solid State Physics,
  Cornell University,
  Ithaca, NY 14853-2501,
  USA
}
\date\today

\begin{abstract}
  We introduce a new toy model for the study of glasses: the hard-matrix model
  (HMM). This may be viewed as a single particle moving on $\mathrm{SO}(N)$,
  where there is a potential proportional to the 1-norm of the matrix. The
  ground states of the model are ``crystals" where all matrix elements have the
  same magnitude. These are the Hadamard matrices when $N$ is divisible by
  four. Just as finding the latter has challenged mathematicians, our model
  fails to find them upon cooling and instead shows all the behaviors that
  characterize physical glasses. With simulations we have located the
  first-order crystallization temperature, the Kauzmann temperature where the
  glass would have the same entropy as the crystal, as well as the standard,
  measurement-time dependent glass transition temperature. Our model also
  brings to light a new kind of elementary excitation special to the glass
  phase: the \textit{rubicon}. In our model these are associated with the
  finite density of matrix elements near zero, the maximum in their
  contribution to the energy. Rubicons enable the system to cross between
  basins without thermal activation, a possibility not much discussed in the
  standard landscape picture. We use these modes to explain the slow dynamics
  in our model and speculate about their role in its quantum extension in the
  context of many-body localization.
\end{abstract}

\maketitle

\section{Introduction}

Much of statistical mechanics is the study of ``toy models,'' minimalistic
distillations of physical systems that capture particular phenomena. The
simplest model of liquids, mono-disperse hard spheres, is also much used as a
model of glassy behavior. In three dimensions, and when compressed rapidly,
this system produces jammed structures with a reproducible packing fraction
\cite{kamien2007random}, but without any obvious order. However, the hard
sphere model falls short in exhibiting all facets of glass phenomenology and is
difficult to treat analytically. We
examine these shortcomings in connection with a new and even simpler model that
might better serve as a model of structural glasses.

The \textit{hard matrix model} (HMM) is a system comprising a single orthogonal
matrix $U\in\mathrm{SO}(N)$ with 1-norm energy:
\begin{equation}\label{eq:hamiltonian}
  \Phi (U)=-\sqrt{N}\;\sum_{i j}|U_{i j}|.
\end{equation}
The matrix elements are not independent, but constrained much like the bond
lengths and angles in a network glass. Their number, $N^2$, is the ``volume" of
the system. The constraints in the hard sphere model are considerably weaker,
and allow small clusters of spheres to act independently when they occur within
a low density fluctuation. In fact, it is precisely such finite sized
equilibrium fluctuations that make the hard sphere system unstable to
nucleating the crystal phase.\footnote{The same mechanism challenges other
packing systems, \textit{e.g.} tetrahedra, as candidate glass models.} If
analogous nucleation events/structures exist for network glasses, they are
poorly understood. The hard matrix model poses this same challenge, in a far
simpler mathematical setting, because it too is potentially unstable to the
analog of crystallization.

Matrix variables have previously been considered in models of glassy behavior without quenched disorder. To the best of our knowledge these have always relied on quartic potentials and fall short of the simplicity of our model (a standard norm imposed on a ubiquitous mathematical object). The square case ($\alpha=1$) of the model studied by Cugliandolo \etal \cite{cugliandolo1995matrix} has only a temperature parameter, like ours, but no glassy properties when the variables are continuous, as in a physical network glass. While it is true that the ground states of model \cite{cugliandolo1995matrix} when restricted to Ising ($\pm 1$) variables coincide with the HMM ground states, we believe the anharmonicities in the continuous variable setting are just as important when modeling glassy systems. The models considered by Solja\u{c}i\'c and Wilcek \cite{soljacic2000minima}, with more parameters, are continuous but the proliferation of $\pm 1$ ground states is enabled by the element-wise anharmonicity $(1-U_{i j}^2)^2$, which defines a rather trivial energy landscape.

In some ways the HMM resembles a mean-field spin glass model: it lacks clear
dimensionality, has ``all-to-all'' interactions, and the global
orthogonality constraint resembles the spherical constraint of the spherical
$p$-spin model \cite{Crisanti_1992_The, Crisanti_1993_The}.  However, unlike most spin glasses the HMM has no quenched
disorder and therefore a sample-independent collection of equilibrium ground
states, which we will see in the next section. At present the precise nature of
its glassy behavior is not well-understood, and the status of hallmark
spin-glass properties, such as the existence of a dynamical transition, is unknown \cite{Castellani_2005_Spin-glass}.  We hope that this
paper inspires investigation into these aspects of the HMM.

In this paper we study several properties of the HMM.
Section~\ref{sec:crystal} describes the ground and low-temperature equilibrium
states of the model, and the transition analogous to crystallization that leads
to them. In Section~\ref{sec:glass} the focus shifts to the main topic: the
metastable glass phase of the model. Section~\ref{sec:quantum} introduces a
quantum generalization, its conjectured phase diagram, and thoughts about
many-body localization. Finally, Section~\ref{sec:conclusions} summarizes our
results, including the many advantages the HMM has over hard spheres.

\section{Equilibrium Thermodynamics}
\label{sec:crystal}

By the generalized mean inequality we know
\begin{equation}
  \Phi (U)\ge -\sqrt{N}N\sqrt{\sum_{i j}|U_{i j}|^2}=-N^2,
\end{equation}
where equality is attained only when the elements of $U$ are equal in
magnitude. The ground states $U^*$ of $\Phi $ are therefore rescaled Hadamard
matrices \cite{hedayat1978hadamard} $U^*=H/\sqrt{N}$, where $H$ has only $\pm
1$ elements, for those $N$ where Hadamard matrices exist. By contrast, the
rigorous ground state characterization of the hard sphere model, the proof of
the Kepler conjecture, required a massive amount of work \cite{hales2005proof}.
There is also a Hadamard matrix conjecture, which asserts that Hadamard
matrices exist for all orders $N$ divisible by four. Empirically, from explicit
enumeration up to $N=32$ \cite{kharaghani2010classification}, the number of
Hadamard matrices $\#(N)$ \cite{OEIS} enjoys robust growth:
\begin{equation}\label{eq:HMentropy}
  \log \#(N)\sim 0.874\, N^{1.6}.
\end{equation}
Ironically, for most (evenly-even) $N$ we lack even a single example
\cite{de2010density}, the smallest open case of the conjecture being $N=668$.
It is for this reason that the hard matrix model still deserves to be called
``hard.'' Simple physics-inspired methods, such as gradient descent on $\Phi
(U)$ from random starting points, almost always fail at finding Hadamard
matrices. The most successful methods for constructing these matrices
\cite{hedayat1978hadamard} are algebraic in nature and require significant
computation. However, because even the most productive of these are based on
first finding sequences of size $N^1$ with special properties, the estimate
(\ref{eq:HMentropy}) suggests that most Hadamard matrices are evading 
discovery.

Not only are the ground states of $\Phi $ known, so are the thermodynamic
equilibrium states in the limit of zero temperature. To see this, parameterize
the neighborhood of a ground state $U^*=H/\sqrt{N}$ with a skew-symmetric
matrix $X$:
\begin{equation}\label{eq:Uparam}
  U(X;H)=H e^X/\sqrt{N}.
\end{equation}
Expanding (\ref{eq:hamiltonian}) for small $X$ we find
\begin{eqnarray}
  \Phi (X;H)&=-\sum_{i j}\mathrm{sgn}(H_{i j})(H(1+X+\frac{1}{2}X^2+\cdots))_{i j}\nonumber\\
  &=-\tr(H^T H(1+X+\frac{1}{2}X^2+\cdots))\nonumber\\
  &=-N^2+\frac{N}{2}\tr(X^T X)+\cdots,\label{eq:localH}
\end{eqnarray}
that is, the potential function reduces to a diagonal quadratic form
independent of the ground state Hadamard $H$. That the contributions to the
free energy are the same for all the ground states is in contrast to the
analogous situation for hard spheres in three dimensions, where the free energy
dependence on the stacking sequence of the close-packed triangular layers was
discovered only recently and required elaborate computations
\cite{mau1999stacking}.

Thanks to the simplicity of the local potentials (\ref{eq:localH}), the
low-temperature limit $\beta\to\infty$ of the HMM partition function
\begin{equation}
  Z(\beta)=\int dU\, e^{-\beta\, \Phi (U)},
\end{equation}
can be evaluated explicitly. With the standard scale convention, the group
invariant measure for small $X$ of the parameterization (\ref{eq:Uparam}) is
\begin{equation}\label{eq:measure}
  dU = \prod_{1\le i<j\le N}\sqrt{2}\; dX_{i j}.
\end{equation}
Since for $\beta\to\infty$ exactly the same Gaussian integral arises around
each Hadamard point of $\mathrm{SO}(N)$, we obtain
\begin{equation}
  Z(\beta)\stackrel[\beta\to\infty]{}{\sim}\#(N)\;e^{\beta N^2}\left(\frac{2\pi}{\beta N}\right)^{N(N-1)/4}
\end{equation}
for those $N$ where $\#(N)>0$.

In the absence of quantum mechanics the entropy has an arbitrary additive
constant and we are free to set $S(0)=0$. This is equivalent to working with
the rescaled partition function $\overline{Z}(\beta)=Z(\beta)/Z(0)$ and
defining the entropy by $S=\log\overline{Z}+\beta\langle\Phi \rangle$, where
$\langle\cdot\rangle$ is the Gibbs average. Using the known volume of
$\mathrm{SO}(N)$ \cite{SOnvolume},
\begin{equation}
  Z(0)=\int dU=2^{(N-1)(N/4+1)}\prod_{k=2}^N\frac{\pi^{k/2}}{\Gamma(k/2)},
\end{equation}
we then have an explicit expression for the HMM entropy in the low temperature
limit. Taking additionally $N$ large, as in a thermodynamic limit, we obtain
the entropy per volume for $\beta\to\infty$ ($e=2.718 \ldots$) :
\begin{equation}\label{eq:entropy}
  s=\frac{1}{N^2}S\stackrel[N\to\infty]{}{\sim}-\frac{1}{4}\log\left(2\beta\sqrt{e} \right)+\frac{1}{N^2}\log \#(N).
\end{equation}
This result looks like it might be used to address the Hadamard matrix
conjecture. By integrating the HMM specific heat $c(\beta)$ in a Markov chain
Monte Carlo (MCMC) simulation, starting at $\beta=0$, it should be possible to
obtain reasonable estimates of $s(\beta)-s(0)=s(\beta)$. A comparison with
(\ref{eq:entropy}), while lacking the precision to determine $\#(N)$ outright,
might still give information about its growth. As we show next, this works for
small $N$. However, large $N$ is inaccessible, and the Hadamard matrix
conjecture remains safe, for the same reasons that make the HMM compelling as a
model for glass.  Cugliandolo \etal \cite{cugliandolo1995matrix} noticed that the $\pm 1$ case of their matrix model has Hadamard (for $\alpha=1$) ground states. However, in the continuous (``spherical") case of this model the ground state is just the $\mathrm{SO}(N)$ manifold and not glassy in the least. Our model has the important advantage that the hard-to-find ground states are also embedded in a non-trivial continuum landscape.

\begin{figure}
  \includegraphics[width=0.9\columnwidth]{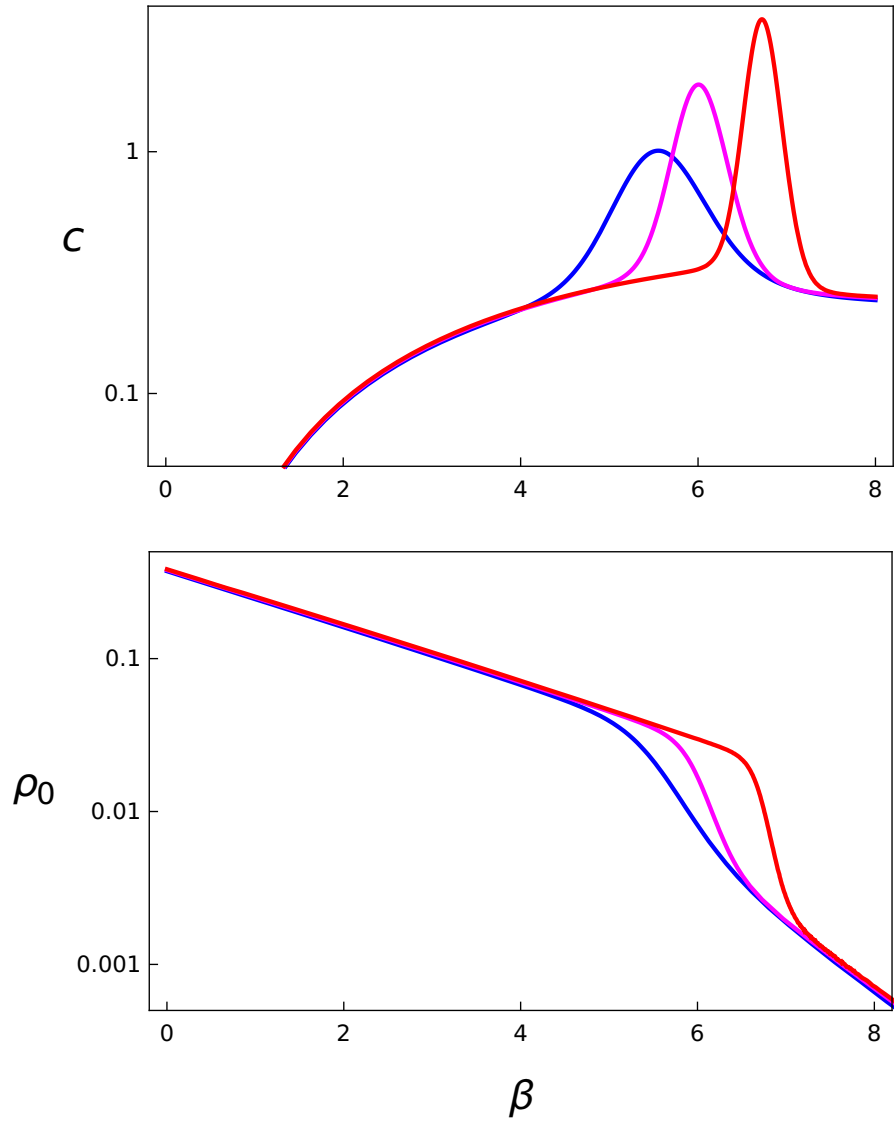}
  \caption{
    Equilibrium specific heat (top) and transition state density (bottom) for
    system sizes $N=12$ (blue), $16$ (magenta), $20$ (red). Narrowing of the specific heat peak and
    abruptness in the drop of $\rho_0$, with increasing $N$, indicate a first
    order transition.  The vertical axis on both plots is logarithmic, and the statistical errors at the sample points are smaller than the thickness of the curves. Comparison of the entropy change defined by the numerical integral of the specific heat agreed with the results of direct Hadamard matrix enumeration to within 0.06\%}.
  \label{fig:fig1}
\end{figure}

Among glass models the HMM is relatively easy to simulate. We sampled the Gibbs
ensemble using MCMC, with elementary transitions generated by Givens rotations
applied to pairs of rows and columns of $U$. The range of the Givens angle was
tuned, at each temperature, so the resulting acceptance rate is 50\%. By
defining a ``sweep" of the system to be rotations attempted on all pairs of
rows and columns, a single MCMC sweep is a reasonable proxy for a time step of
true dynamics, since the number of actual moves per sweep scales with the
number of continuous degrees of freedom. MCMC simulation of our model's
equilibrium thermodynamics benefits from the use of parallel tempering, and we
use this method when the stable equilibrium is achievable and simulation time
scales are not our interest. When parallel tempering was used, as for the
results of Figure~\ref{fig:fig1}, temperatures were selected to optimize the
replica transit time \cite{Katzgraber_2006_Feedback-optimized}.

Figure \ref{fig:fig1} shows the MCMC specific heat, or heat capacity per unit
volume, $c=C/N^2$, for $N=12, 16, 20$. The evidence for a first-order phase
transition in the infinite system is strong. We will say the system at high
temperatures is in the \textit{liquid phase}, and at low temperatures in the
\textit{Hadamard phase}. The latter is in fact a collection of $\#(N)$ phases,
each associated with a different Hadamard ``crystal" ground state. At $N=20$ we
are already at the limit of being able to maintain thermal equilibrium with
Givens rotations. Near the specific heat peak, MCMC simulations require
$2\times 10^9$ sweeps per measurement, while precise measurement by parallel
tempering took $1\times 10^8$ sweeps per temperature with $2\times10^7$
adjacent replica steps. A good test of the accuracy of the $c(\beta)$ curve is
the corresponding entropy integral. When compared against (\ref{eq:entropy}),
this reproduced the known Hadamard count $\#(20)\approx 2\times 10^{45}$
\cite{OEIS} to within a factor of 1.28, a 0.06\% error in the
entropy.\footnote{This is before making large $N$ approximations, as the
Hadamard count makes a subextensive contribution. The counts in OEIS A206711,
which include Hadamards of negative determinant, were divided by two.} The low
temperature limit $c=1/4$ is simply the equipartition value $1/2$ for quadratic
potentials combined with the number of continuous modes being only half the
system volume.

The HMM challenges our understanding of first order phase transitions. Consider
the notion of phase coexistence. We suspect that the HMM does not exhibit phase
coexistence in the usual sense. Suppose the system is prepared with energy
density halfway between that of the pure phases at the transition, say by MCMC
sampling at the temperature of the specific heat peak (in a modest sized system
where this is possible). What might such a system look like? We doubt that the
configuration will be mixed-phase in the usual sense, say a proper Hadamard
submatrix within a ``liquid matrix." If such mixed configurations existed, with
continuously variable composition, then it would also be possible to have
critical nuclei for crystallization, contrary to the extreme degree of
metastability we observe already for $N=24$.

While all of our numerical experiments used MCMC, true dynamics could be
simulated by time-evolving the unconstrained system (see Appendix)
\begin{eqnarray}
  \mu N(\ddot{U}+\dot{U}\dot{U}^T U)&=-\frac{1}{2}\left(\nabla\Phi-U(\nabla\Phi)^TU\right)\label{eq:EOM}\\
  &=\frac{\sqrt{N}}{2}\left(\mathrm{sgn}(U)-U \mathrm{sgn}(U)^TU\right),\label{eq:force}
\end{eqnarray}
with initial constraints
\begin{equation}
  U(0)U^T(0)=1,\quad \dot{U}(0)U^T(0)+U(0)\dot{U}^T(0)=0.
\end{equation}
The left-hand side of (\ref{eq:EOM}) generates free motion on $\mathrm{SO}(N)$,
and the scaling of the mass with $N$ was chosen so the equations for small
oscillation about the Hadamard minima,
\begin{equation}
  \mu\ddot{X}=-X,
\end{equation}
are independent of $N$. From (\ref{eq:force}) we see that the mechanical
equilibrium points of $\Phi (U)$ correspond to orthogonal matrices with the
following symmetry property:
\begin{equation}\label{eq:symprop}
  U^T\mathrm{sgn}(U)= \mathrm{sgn}(U)^T U.
\end{equation}
These are a superset of the Hadamard matrices and it is their high abundance
that defeats the prospect of finding Hadamard matrices by gradient descent on
$\Phi (U)$. It is tempting to look at property (\ref{eq:symprop}) as a set of
geometrical constraints of exactly the right number to fix all the continuous
variables of an orthogonal matrix, in analogy with isostaticity in jammed
sphere packings or rigidity of ball-and-stick network models. While this
perspective can be useful for identifying good glass formers when constituents
are modeled geometrically \cite{phillips1979topology}, in our case it is simply
an automatic consequence of a sufficiently well-behaved potential function.

\begin{figure}
  \includegraphics[width=\columnwidth]{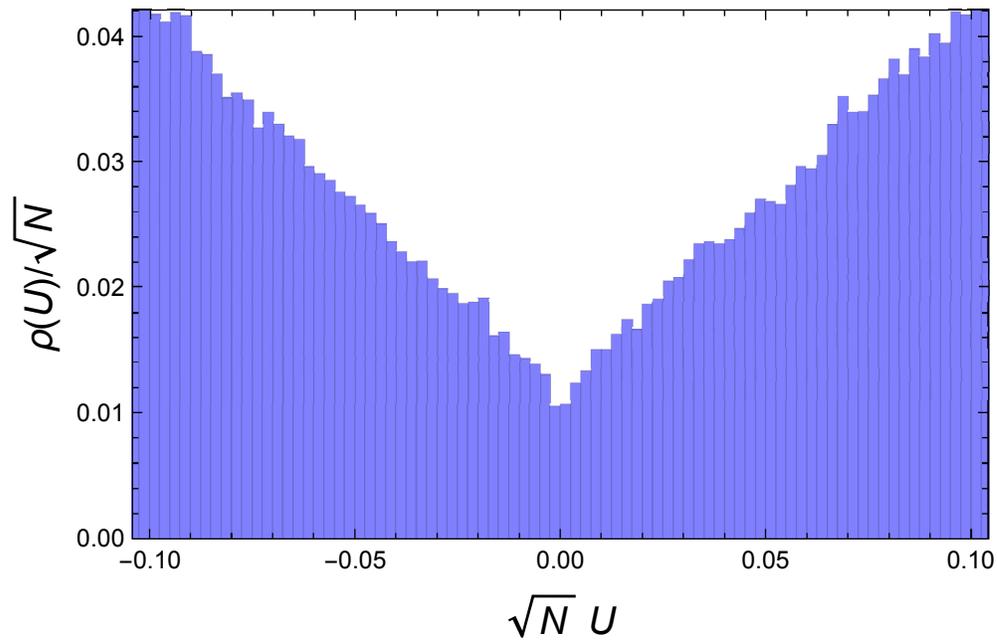}
  \caption{
    Detail of the distribution of matrix elements, near $U=0$, of equilibrium
    points generated by gradient descent. All equilibria are fragile in the
    sense that some fraction of the matrix elements are near a ``transition
    state" in their contribution to the energy.
  }
  \label{fig:fig2}
\end{figure}

The mechanical equilibrium points are relevant for the dynamics at low
temperature. Figure \ref{fig:fig2} shows a detail of $\rho(U)$, the
distribution of the individual matrix elements, near $U=0$ where their
contribution to the energy is highest. This distribution was generated by
gradient descent from random points on $\mathrm{SO}(32)$. The property
$\rho(0)>0$, which seems to hold in the thermodynamic limit, confers a
fragility to the mechanical equilibria. Consider the set of matrix elements
whose values are within some fixed, small distance of zero. For each of these
there is a small geodesic motion that brings the matrix element to the
transition state point of its energy, the cusp at $U=0$. These single-element
transition states are likely also transition states for the system as a whole
because the regular contribution to $\Phi$ (from the other matrix elements)
only changes quadratically and the motion is small. Unlike phonon modes, for
which the perturbation sees an opposing/restoring force, the opposite is true
when a matrix element crosses $U=0$. These \textit{rubicon modes}, present at
all temperatures with density proportional $\rho(0)$, provide a mechanism for
the system to sample the energy landscape without ever having to surmount an
energy barrier. As we will see, the growth of the dynamical mixing time at low
temperatures may not be the result of activated processes at all, but a byproduct
of a sharply reduced rubicon density.

\begin{figure}
  \includegraphics[width=\columnwidth]{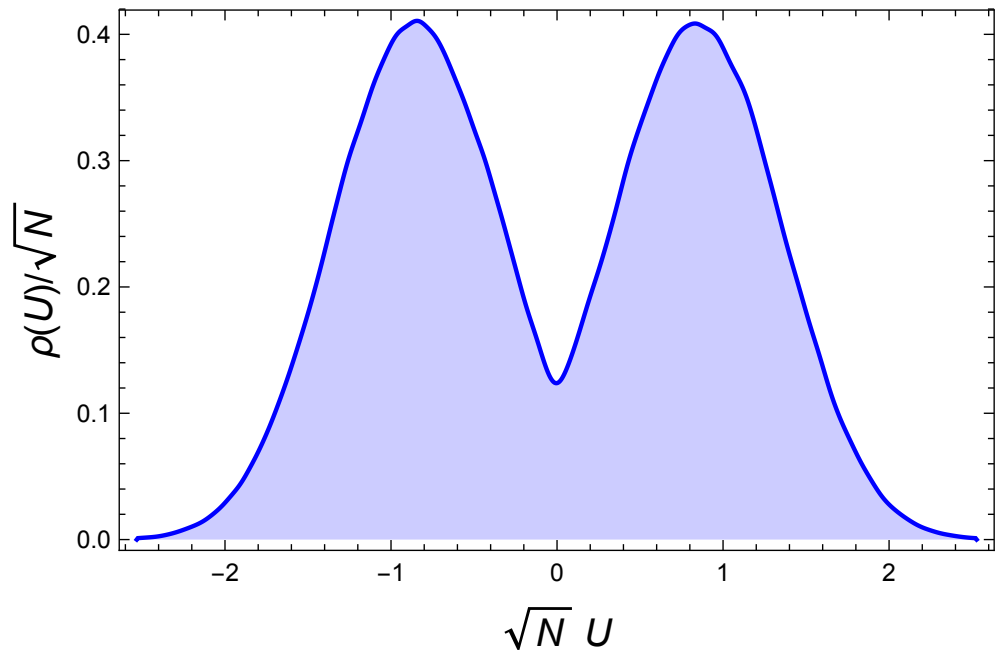}
  \caption{
    Probability distribution of the matrix elements at $\beta=3$, from a $N=32$
    simulation, shows the liquid is strongly correlated at temperatures well
    above the phase transition.
  }
  \label{fig:fig3}
\end{figure}

We believe the \textit{transition state density} $\rho_0=\rho(0)/\sqrt{N}$ is
the key structural property of our model. It is known that $\rho(U)$ is
normally distributed at $\beta=0$ in the large $N$ limit
\cite{diaconis2003patterns}, and yet as shown in Figure \ref{fig:fig3}, it is
strongly bimodal already at $\beta=3$. As in physical glass formers, the liquid
phase of the HMM is strongly correlated at temperatures well above
crystallization, and the smallness of $\rho_0$ is a useful measure of this. The
first-order nature of the liquid/Hadamard phase transition is clearly seen in
the discontinuity-tending behavior of $\rho_0$, with increasing $N$, shown in
Figure~\ref{fig:fig1} for the same system sizes discussed earlier. Finally, by
being linked to dynamics via fragile equilibria and rubicon modes, $\rho_0$
will also be relevant to the discussion of glasses to which we turn next. In
brief, we find $\rho_0$ has a simple Arrhenius behavior in the glass, much like
the concentration of defects in a crystal. The onset of slow equilibration in
glasses might therefore be less a consequence of rising energy barriers and
more the result of a scarcity of modes that can stir the system. In the
Hadamard phase $\rho_0\sim\sqrt{\beta/\pi}\,e^{-\beta}$ also has the Arrhenius
form, but with a steeper slope than in the glass. This quantity is not linked
to transition states (rubicons) in the Hadamard phase because, unlike the
fragile equilibria, it becomes small simply by the suppression of the large
thermal fluctuations required to create zero matrix elements.

\section{Metastable glass}
\label{sec:glass}

\begin{figure}[htpb]
  \includegraphics[width=\columnwidth]{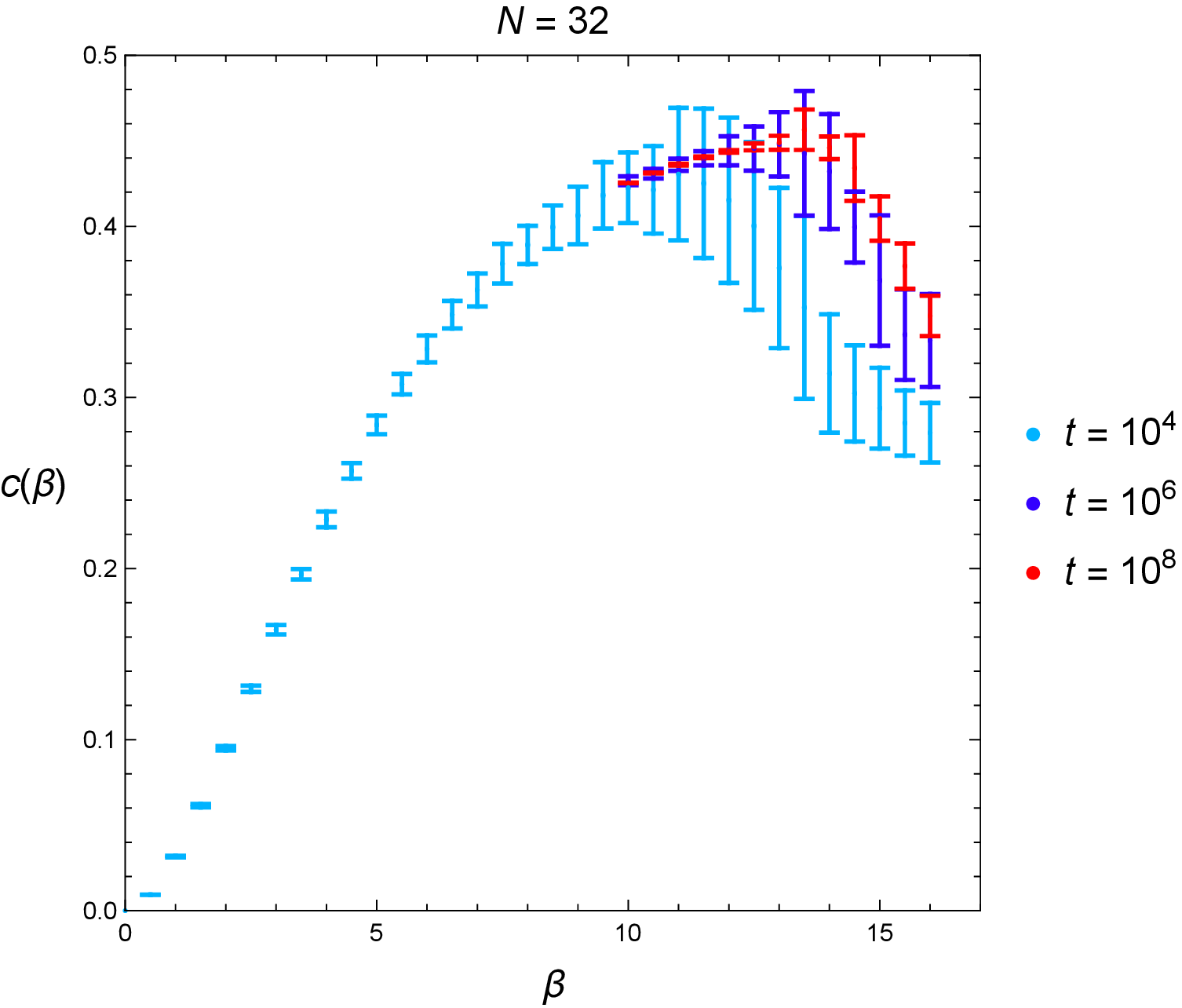}
  \caption{
    Specific heat of the $N=32$ system with increasing number of MCMC sweeps
    $t$ per measurement. The error bars show one standard deviation in the
    measurements of 20 systems.
  }
  \label{fig:fig4}
\end{figure}

\begin{figure}
  \includegraphics[width=\columnwidth]{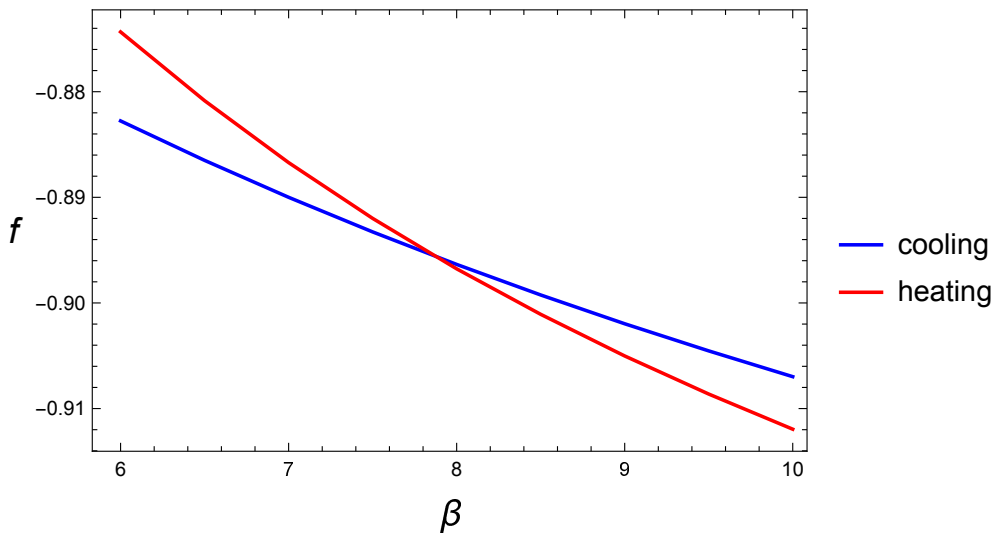}
  \caption{
    MCMC free energies of the $N=32$ system cooled from the liquid and heated
    from one of the crystal phases.
  }
  \label{fig:fig5}
\end{figure}

Much of glass phenomenology
\cite{ediger1996supercooled,debenedetti2001supercooled} is captured in the
series of specific heat measurements shown in Figure \ref{fig:fig4} for the
$N=32$ system. These were initialized in the liquid phase and show no sign of
an anomaly at the expected crystallization temperature, even at the longest
equilibration times, or number of sweeps $t$. We estimated $\beta_H$ for $N=32$ by measuring the
free energies of the metastable liquid and Hadamard phases by cooling the
former and heating the latter. The free energy crossing is shown in Figure
\ref{fig:fig5} and locates the transition at $\beta_H=7.9$. Instead of a peak
in the specific heat at $\beta_H$, the cooled $N=32$ liquid exhibits a gently
rising $c(\beta)$, with each reduction in cooling rate hopefully revealing a
better equilibrated metastable phase: the HMM \textit{glass}.

Upon closer examination of Figure \ref{fig:fig4} we see that there are actually
two distinct phenomena. First, the fact that the drop in $c(\beta)$ shows signs
of recovering with increasing $t$ can be interpreted by the system falling out
of equilibrium: on the low temperature side of the broad specific heat peak the
equilibration time is insufficient for the system to sample the full,
equilibrium range of energy fluctuations. The second phenomenon concerns the
error bars. These show one standard deviation in the measurements of 20
systems. We see a relatively abrupt transition between a regime where
increasing $t$ has the desired $\sim 1/\sqrt{t}$ convergence to a unique
average, to another regime, at low temperature, where increasing $t$ has very
little effect and instead it is the systems that have become unique.

\begin{figure}[htpb]
  \includegraphics[width=\columnwidth]{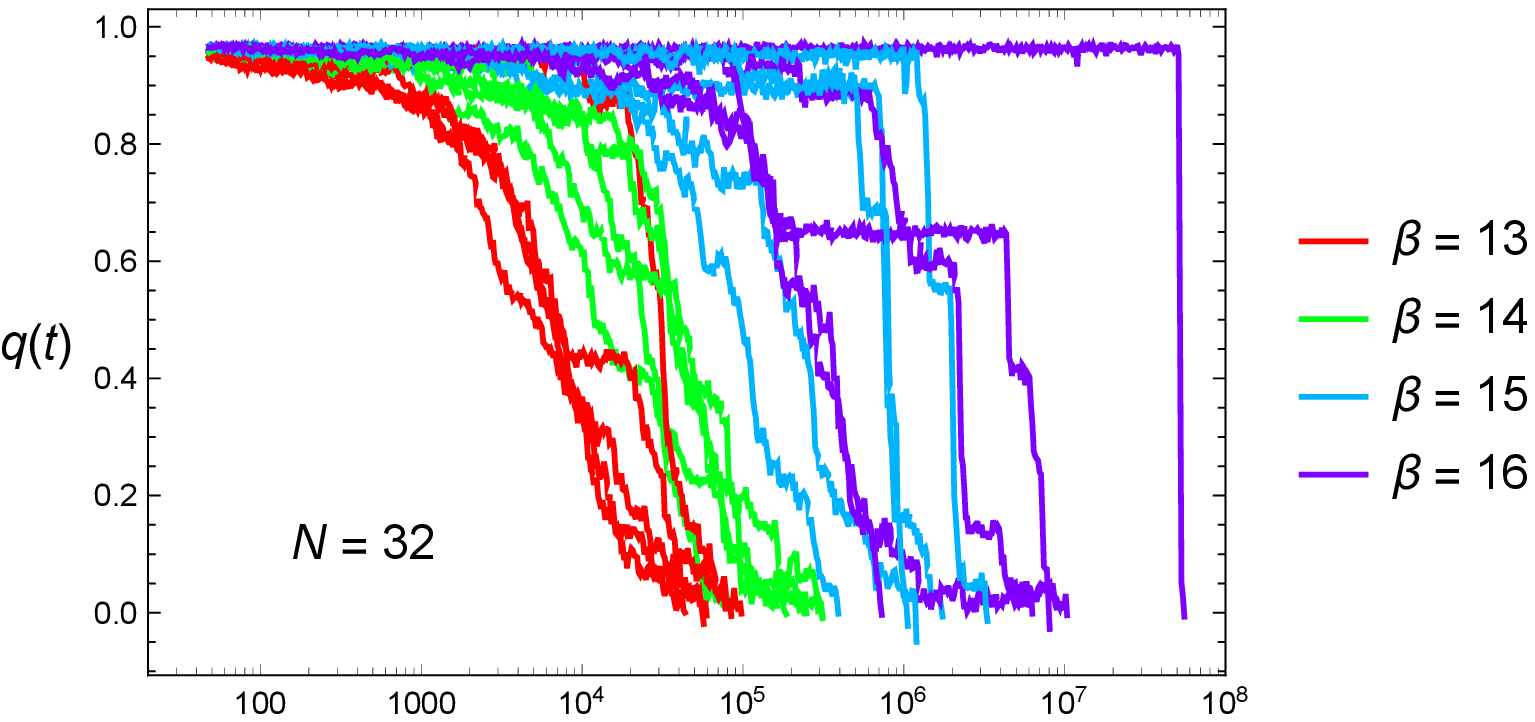}
  \includegraphics[width=\columnwidth]{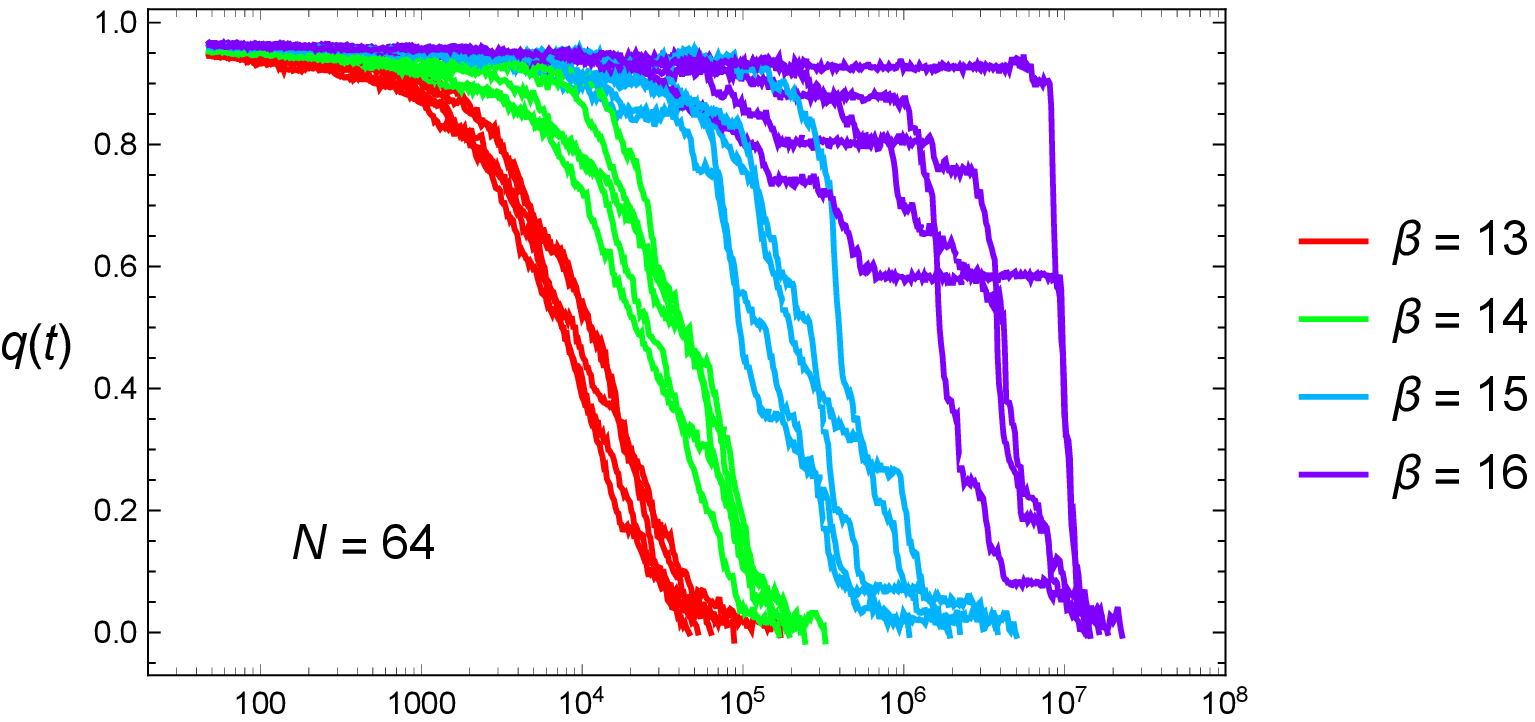}
  \includegraphics[width=\columnwidth]{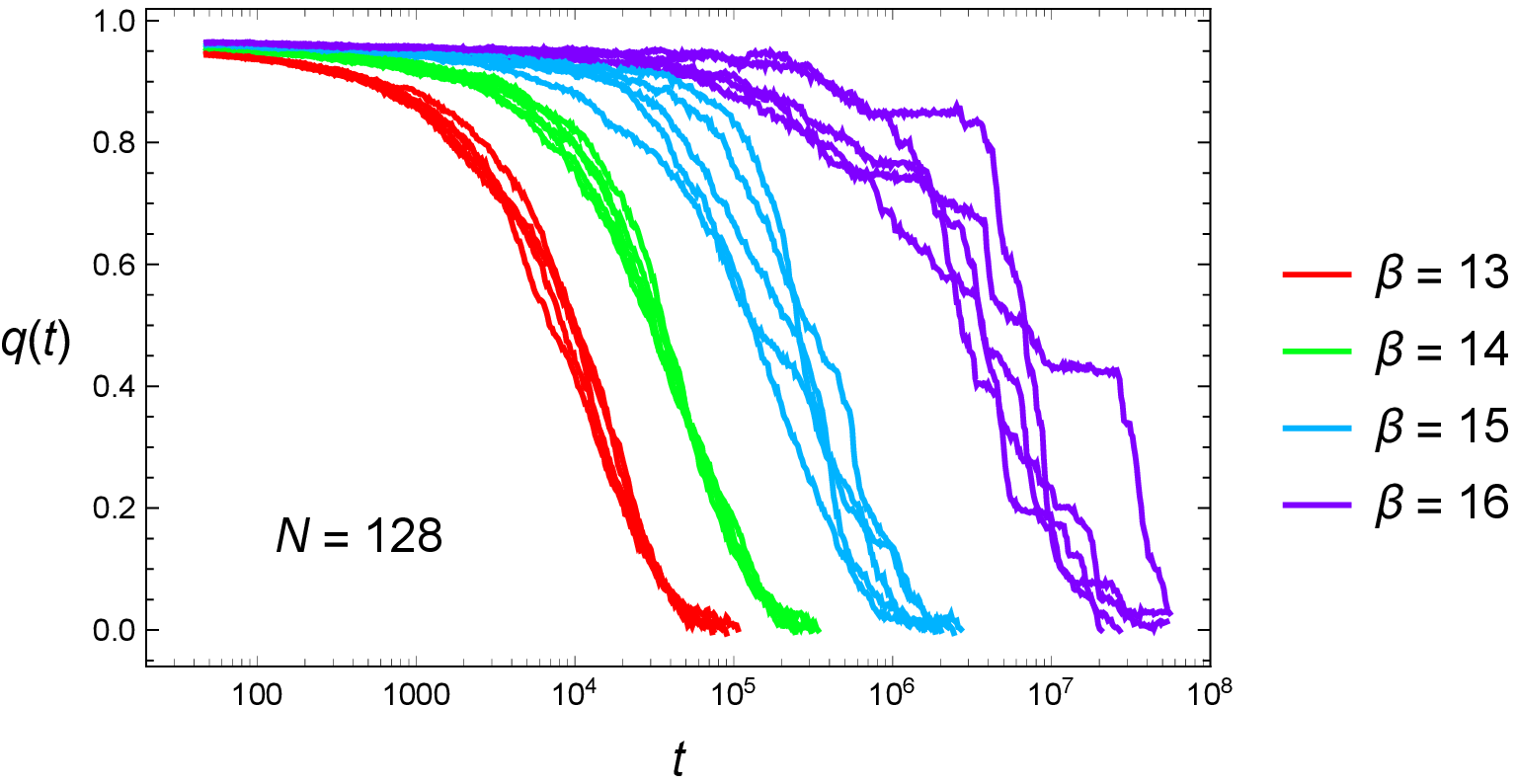}
  \caption{
    Time evolution of the correlation (\ref{q(t)}) for different inverse
    temperatures and system sizes. Equal increments in $\beta$ translates
    $q(t)$ along the logarithmic time axis by increasing amounts.
    Reproducibility of the corresponding time-rescaling symmetry, over the five
    experiments at each $\beta$, holds only when the system size is
    sufficiently large.
  }
  \label{fig:fig6}
\end{figure}

The two phenomena we see in the glass phase specific heat are disentangled by
examining the time evolution of the correlation
\begin{equation}\label{q(t)}
  q(t)=\frac{1}{N}\tr U^T(0)\,U(t).
\end{equation}
As in the specific heat measurements, one unit of the time $t$ is a single MCMC
sweep (with Givens rotations tuned to have 50\% acceptance rate), and $10^6$
sweeps were used to equilibrate $U(0)$. Figure \ref{fig:fig6} shows the decay
of $q(t)$ with time evolution, for four temperatures and three system sizes.
Each experiment was repeated five times.

The results for the largest system size, $N=128$, are consistent with there
being no dynamical transition. As $\beta$ increases, the correlation $q(t)$ has
the same decay after a suitable rescaling of the time. However, the results
also show that one might be led to a different conclusion when the
thermodynamic limit ($N\to\infty$) is not taken into consideration. For
example, just based on the results for $N=64$ one might conclude there is a
transition to a phase with ``persistent history" for $\beta$ greater than about
14.

\begin{figure}
  \includegraphics[width=\columnwidth]{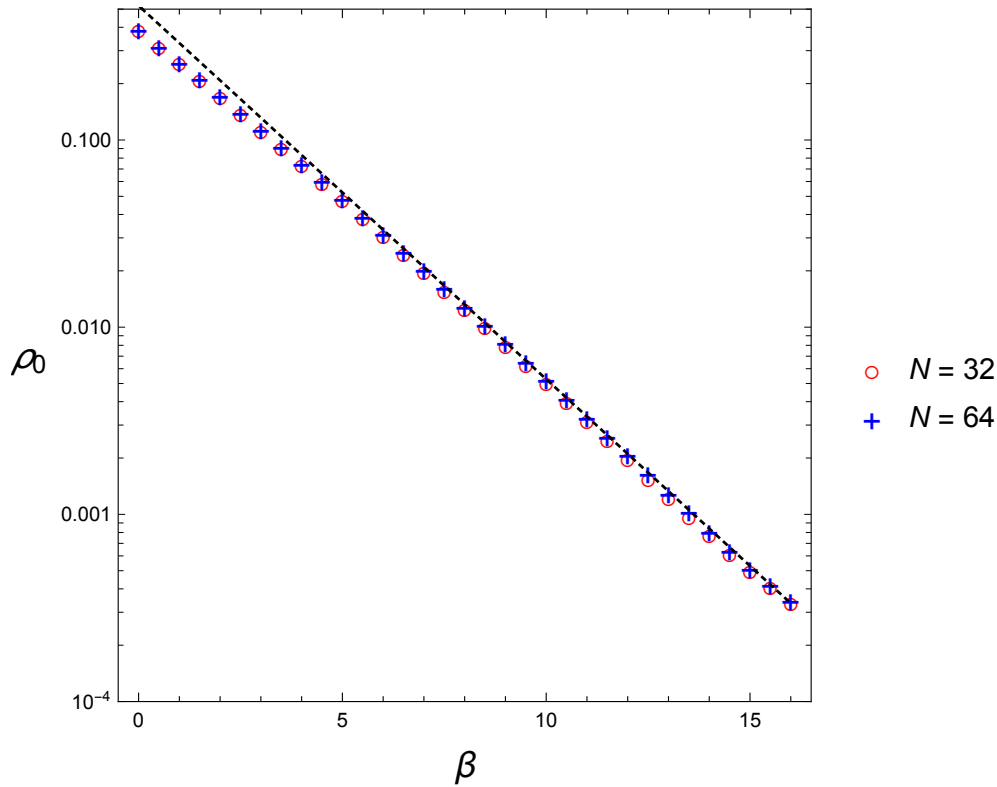}
  \caption{
    Arrhenius behavior of the transition state density $\rho_0$ for two system
    sizes. Shown are results for averages over $10^6$ sweeps; results with only
    $10^4$ sweeps are practically indistinguishable. The dashed line is a
    linear fit at the low temperature end.
  }
  \label{fig:fig7}
\end{figure}

The onset, at large $\beta$, of dynamics dictated by events in the distant
past, can be explained using the transition state density $\rho_0$. We first
observe, as shown in Figure \ref{fig:fig7}, that $\rho_0$ has a simple
Arrhenius thermal behavior:
\begin{equation}\label{eq:rhobeta}
  \rho_0(\beta)\propto e^{-\beta e_0},
\end{equation}
where $e_0\approx 0.46$. For both of the system sizes shown, the range in
values from 20 experiments is smaller than the plot symbols already at $10^4$
sweeps and about 10 times smaller than this when the average is over $10^6$
sweeps. The plot also confirms that the density $\rho(U)$, with $U$ scaled by
$\sqrt{N}$, is a thermodynamic density, independent of $N$, all the way to the
$U=0$ transition-state elements. At the same time, the Arrhenius behavior of
$\rho_0$ challenges the achievability of the thermodynamic limit in our model.
If we assume that rubicon modes are the dominant mixing mechanism, then we
should expect equilibrium in a thermodynamic sense only if the total number of
rubicon modes is $O(N^2)$, and a breakdown occurs when $N^2\rho_0(\beta)\sim c$
for some $c=o(N^2)$. For example, taking $c=10$ and using (\ref{eq:rhobeta}),
we should expect thermodynamic-limit behavior only for $\beta$ below 8.7, 11.7
and 14.7, respectively, in the $N=32,64,128$ systems.

We now turn to the problem of estimating the mixing time $\tau(\beta)$, that
is, the time-rescaling which collapses the the correlations $q(t)$ onto a
single curve. As we argued above, this is an academic exercise in asymptotics
since the result may hold only when the system volume is allowed to grow
exponentially with inverse temperature. Our derivation makes a number of leaps
in logic and is mostly an attempt to interpret an asymptotic form that
empirically collapses the data despite having just a single parameter.

\begin{figure}
  \includegraphics[width=\columnwidth]{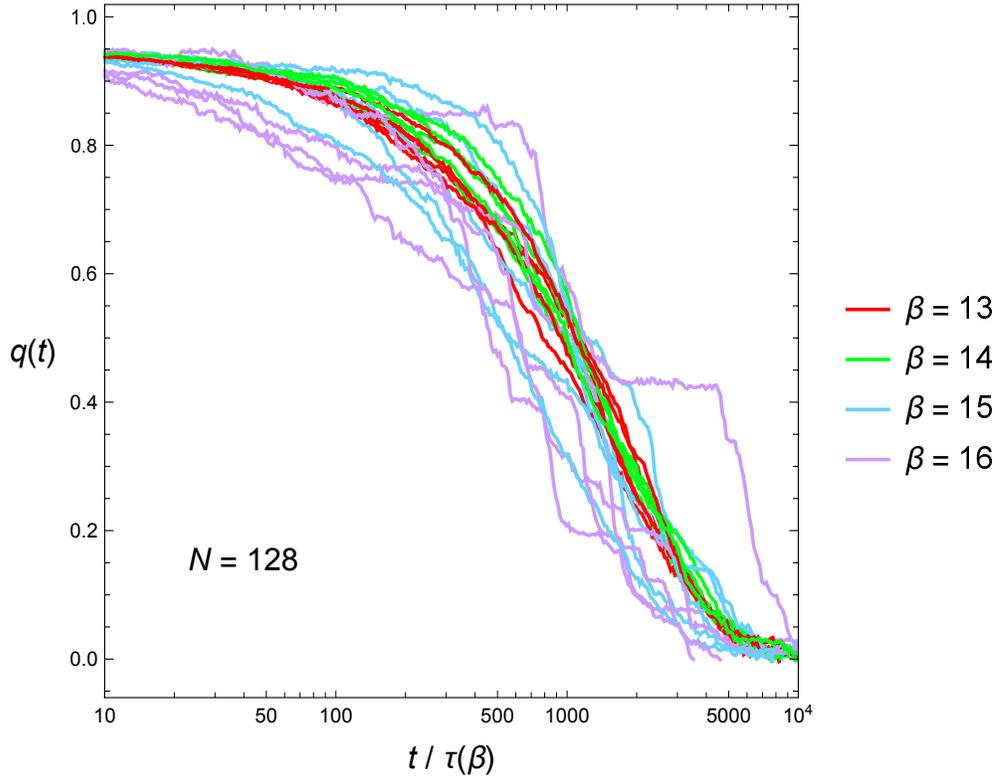}
  \caption{
    Collapse of the $q(t)$ curves in Figure \ref{fig:fig6}, for $N=128$, upon
    scaling the time by the mixing time (\ref{eq:taubeta}) with $b=5.2$.
  }
  \label{fig:fig8}
\end{figure}

The main ingredient of the estimate is the idea that the number of independent
modes in the mixing dynamics is proportional to the number of rubicons in the
system, $\rho_0 N^2$. Alternatively, we can think of the accessible states as
lying in a space of dimension
\begin{equation}\label{eq:channeldim}
  D(\beta)\propto \rho_0(\beta) N^2
\end{equation}
within $\mathrm{SO}(N)$ and on which there is free motion. If $\xi(\beta)$ is
the distance scale (on $\mathrm{SO}(N)$) associated with each rubicon mode,
then a sufficient condition for mixing is the volume-of-states criterion
\begin{equation}\label{eq:volsampling}
  \xi(\beta)^{D(\beta)}\sim e^{S_0},
\end{equation}
where $S_0=O(N^2)$ is an assumed temperature-independent entropy of the glass
(conventionally designated ``configurational"). The nature of the glassy
dynamics, as expressed in this relation, is that the rubicon mode amplitudes
$\xi$ are forced to be very long as the dimension $D$ becomes small at large
$\beta$. Assuming diffusive rubicon motion, the mixing time enters by the
relation
\begin{equation}\label{eq:diff}
  \xi(\beta)\propto \sqrt{\tau(\beta)}.
\end{equation}
Combining (\ref{eq:rhobeta}), (\ref{eq:channeldim}), (\ref{eq:volsampling}) and
(\ref{eq:diff}) results in the large-$\beta$ behavior
\begin{equation}\label{eq:taubeta}
  \tau(\beta)\propto \exp\left(\exp(\beta e_0-b)\right)
\end{equation}
with a single undetermined parameter $b$. The collapse of the $q(t)$ curves for
the $N=128$ system is shown in Figure \ref{fig:fig8} for $b=5.2$ and with 1 as
the proportionality factor in (\ref{eq:taubeta}).

If we fix a proportionality factor $a$ in (\ref{eq:taubeta}) so that
$q(\tau)=1/2$ and designate this $\tau$ as the ``measurement time" in an
experiment, then solving (\ref{eq:taubeta}) for $\beta$ defines the glass
transition (inverse) temperature:
\begin{equation}
  \beta_g(\tau)=\left(\log(\log(\tau/a))+b\right)/e_0.
\end{equation}
The parameter $a$ depends on the details of the dynamical equations. For our
50\%-acceptance-Givens-rotation dynamics, $a\approx 1000$.

After the exercise of analysing the behavior of the correlations $q(t)$, in
time as well as system size, we are in a better position to interpret the
specific heat results of Figure \ref{fig:fig4}. The default interpretation is
that by the application of extraordinary resources, not just in measurement
time but also system size, the specific heat will reveal itself to saturate at
some value $c_g$ greater than the $c=1/4$ from the harmonic vibrations in the
Hadamard phase. This predicts that the entropy of the glass at low temperatures
has the form
\begin{equation}\label{eq:glassentropy}
  s_g(\beta)=-c_g\log (\beta/\beta_0),
\end{equation}
where $c_g\approx 0.46$ from the $N=32$ simulation and $\beta_0\approx 3.63$
comes from integrating the non-constant part of the specific heat. Comparing
(\ref{eq:glassentropy}) with (\ref{eq:entropy}) we find that the Hadamard
(``crystal") entropy would exceed the glass entropy for $\beta>\beta_K\approx
26$. This state of affairs is known as the Kauzmann paradox
\cite{kauzmann1948nature} and the $\beta_K$ so defined is the inverse Kauzmann
temperature. If we take our measurement time estimate above seriously, then to
sample such a paradoxical glass configuration a simulation would require of
order $10^{400}$ time steps. We note that the abundance of Hadamard
matrices---each a distinct ``crystal'' phase---does not resolve the paradox
because this contribution to the entropy is non-extensive.

To resolve the Kauzmann paradox in our system one should not overlook the
possibility that the specific heat has a maximum, even when the system is fully
equilibrated. Provided $c(\beta)$ has a suitable approach to the Hadamard value
$c=1/4$ at large $\beta$, the glass entropy can remain above that of the
ordered Hadamard phase. The only thing odd about this scenario would be the
somewhat large $\beta$ of the specific heat maximum. But even this can be
dismissed because the known transition to the stable low temperature phase,
$\beta_H$, is also large.

On a more fundamental level one might ask whether there is even a paradox in
need of resolution. The paradox is usually presented from the perspective of
``landscape theory" \cite{debenedetti2001supercooled}, where the system is seen
as sampling a collection of local minima in a complex potential energy
landscape. In this picture the entropy has a configurational contribution that
counts the number of accessible minima, and a vibrational contribution
associated with the harmonic modes within a representative local minimum. If
one makes the reasonable assumption that the vibrational contributions in the
glass and crystal are the same, one arrives at the paradox that the glass has
fewer configurational states than the crystal. However, we next argue that this
partitioning of the entropy is naive and may even be logically inconsistent.

For the ``landscape" partitioning of the entropy to be valid, surely the
specific heat must be close to the vibrational value ($1/4$ in our model) at
inverse temperatures above $\beta_K$. This value is independent of the strength
of the harmonic restoring forces at the local minimum. To claim any other value
undercuts the argument that the only relevant modes are vibrations. If one
invokes activated ``minimum-hopping" to explain the enhanced specific heat
(energy fluctuations), then these should be included in the tally of relevant
modes along with vibrations. On the other hand, it seems inconsistent to us to
invoke such specific heat-boosting modes when cooling to the paradoxical point,
and then neglecting them when accounting for the resulting entropy. 

The HMM can be helpful in clarifying the mode analysis of the glassy state
because we believe we have identified the modes most relevant at low
temperatures. These are the rubicons, that is, the elementary transition-state
modes whereby the system ``crosses" from one basin to another. These modes
clearly play a role in boosting energy fluctuations and raising the specific
heat above the vibrational value. At the same time, because these
non-vibrational modes persist to arbitrarily low temperature, it is incorrect
to explain the limiting entropy strictly in terms of vibrations. In fact, we
see no \emph{a priori} reason why exotic modes could not have the net effect of
bringing the glass entropy below that of the crystal at low temperatures.

It seems reasonable to expect that rubicon modes exist in other models, though
perhaps not as transparently. Characterized most broadly, rubicons (i) are
elementary modes associated with transition states, (ii) are not thermally
activated, and (iii) have a finite density even when the system is quenched to
zero temperature. The third property guides us in discovering rubicons in a new
system. After a quench one checks whether the system is almost always poised
very near a transition state in the energy. It is interesting that empirically
the rubicon density in the HMM has an Arrhenius thermal dependence. We do not
know whether to expect this to be a general property as well, and in fact we do
not understand how this simple behavior arises in the HMM.

How would rubicons manifest themselves in a real network glass? We address this
from the perspective of an all-atom molecular dynamics simulator, as this kind
of `experiment' affords the most direct access to the physical processes.
Suppose the researcher has prepared the system at temperature $T$ by some
cooling protocol, quenches the kinetic energy, and relaxes the system to a
configuration of mechanical equilibrium. In the landscape picture, without
rubicons, these mechanical equilibria are unremarkable and only weakly
sensitive to $T$. The slow dynamics, if kinetic energy were reintroduced, comes
about through Arrhenius activation over barriers, as energy gets redistributed
through weakly anharmonic processes. With rubicons this picture would be quite
different. The mechanical equilibria (after quenching/relaxing) would always be
fragile, that is, there would be a finite density of modes (rubicons) with
excitation energy $\epsilon$ to a transition state, in the limit
$\epsilon\to 0$. With rubicons around, the system always `flows', but the rate
decreases because the rubicon density is strongly temperature dependent. The
last point is important, as our simulations of the HMM brought to light: When
the rubicon density is so low that their mean number is less than $O(1)$, the
glass will not be in equilibrium. The greatest challenge in realizing
equilibrium in all-atom network glass simulations, in the rubicon picture, will
therefore be the very large system sizes required, and not so much the long
time scales of activated processes.

\section{Quantum model}
\label{sec:quantum}

To better address the Kauzmann conundrum in the HMM, while also giving a
rigorous definition of states and dynamics, we introduce a quantum extension of
the model. Whereas the HMM had no parameters, the Hamiltonian of the quantum
model now has one:
\begin{equation}\label{eq:qmodel}
  \mathcal{H}=-\frac{1}{4N\mu}\Delta_U+\Phi(U).
\end{equation}
Here $\Delta_U$ is the Laplace-Beltrami operator on $\mathrm{SO}(N)$ and the
mass $\mu$ is the sole parameter. Setting the scale factor of the Laplacian so
that locally ($U=e^X$, $X$ small)
\begin{equation}\label{eq:laplacian}
  \Delta_U=\sum_{1\le i<j\le N}\frac{\partial^2}{\partial X_{i j}^2},
\end{equation}
we see, using (\ref{eq:localH}), that the frequency of harmonic oscillation
about the Hadamard minima is $\omega=1/\sqrt{\mu}$. At fixed $\beta$, taking
the limit $\mu\to \infty$ so that $\beta (\hbar\, \omega)\to 0$, we recover the
classical HMM which has a first-order, thermal transition at $\beta=\beta_H$.
In the $\beta$-$\mu$ plane we expect this to become a line of first-order
transitions. Since $\mu\to 0$ corresponds to a free particle on
$\mathrm{SO}(N)$, a model with no thermal transition, the simplest scenario for
the interior of the phase diagram is that the line of first-order transitions
terminates on the zero-temperature axis as sketched in Figure \ref{fig:fig8}.
The endpoint of the phase boundary, at $\mu=\mu_c$, would then be a quantum
(zero-temperature) phase transition. Along the phase boundary the transition
state density $\rho_0$ drops discontinuously upon crossing into the Hadamard
phase.

\begin{figure}
  \includegraphics[width=\columnwidth]{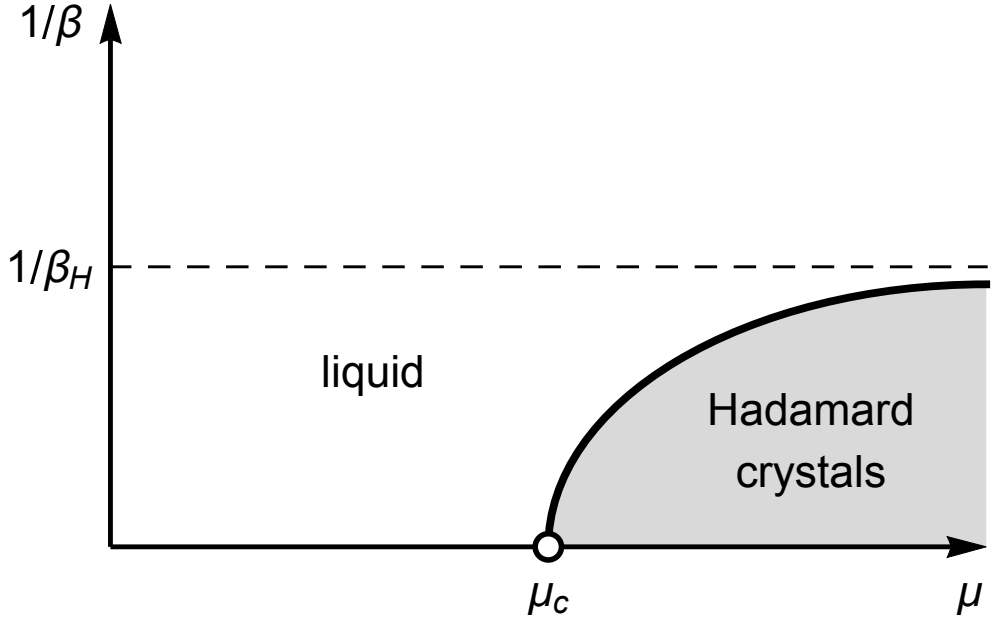}
  \caption{Conjectured equilibrium phase diagram of the quantum hard matrix model.}
  \label{fig:fig9}
\end{figure}

For investigating quantum equilibrium behavior, the hard matrix model has
another advantage over hard spheres in that simulations are a straightforward
extension of the classical case. In the standard path-integral scheme, for
inverse temperature or imaginary ``time" $\beta$ divided into increments
$\Delta \beta$, there will be $\beta/\Delta \beta$ classical ensembles, each
subject to the potential (\ref{eq:hamiltonian}). The only new feature in the
quantum simulation is that there is a kinetic contribution,
\begin{equation}\label{eq:Ucorr}
  \tr U^T(\beta)\,U(\beta'),
\end{equation}
for adjacent imaginary times. Sampling the quantum ensemble could still be
implemented with bounded-range Givens rotations, now including coupling terms
between the two matrices at adjacent times. By contrast, MCMC updates for the
quantum hard sphere model require complex world-line reconnections
\cite{ceperley1986path} to impose the permutation symmetry of the spheres.

The liquid and Hadamard phases acquire new interpretations when we restrict to
the zero temperature axis. In the Hadamard phase the configurations ($U$) are
localized at one of the Hadamard points of $\mathrm{SO}(N)$. In the limit
$\mu\to \infty$ the ladder of excitations becomes more perfectly harmonic and
the wave functions have Gaussian decay away from these points. On the other
side of the quantum phase transition, $\mu<\mu_c$, the matrix $U$ is quantum
delocalized.

The onset of localization (at a Hadamard point) for $\mu\to\mu_c^-$ can be
detected by measuring the expectation of (\ref{eq:Ucorr}), not just for
adjacent times, but for $|\beta-\beta'|$ that span the full range of times in
the simulation. While establishing a new quantum phase transitions is a
worthwhile end in itself, we should not lose track of the glass phenomenology
that motivated the HMM in the first place. In particular, since the
$\mu\to\infty$ limit of the line of first order transitions in Figure
\ref{fig:fig9} is practically non-existent for even moderately large $N$, one
may reasonably expect the entire diagram is covered by a single
liquid/metastable-glass phase. It is in this setting we should ask what new
insights the quantum-HMM can provide.

One way to ask whether quantum mechanics is at all relevant is to compare the
order of the limits $\mu\to\infty$ and $\beta\to\infty$. On taking the former
limit first we get the classical model, and in the previous section we learned
that this case brings no surprises. The classical model does not appear to have
a dynamical transition: the mixing time grows longer indefinitely, all the way
to zero temperature.

The opposite order of limits, where $\mu$ is fixed as we go to zero
temperature, brings to mind the phenomenon of Anderson localization
\cite{anderson1958absence}. For this limit we are interested in the low-lying
quantum states of a quantum particle moving on $\mathrm{SO}(N)$ in the presence
of a potential that may ``look" as random to the particle as an Anderson model.
However, before we explore this part of phenomenology we should heed the lesson
of many-body localization (MBL) \cite{nandkishore2015many} that the temperature
of the system, when in isolation, may not be well defined. We therefore need to
look for another way to parameterize the transition state density $\rho_0$,
which, as before, is the core property used by our analysis. As a substitute we
use the excitation energy density
\begin{equation}
  \epsilon=\Phi(U^*)/N^2+1
\end{equation}
of the fragile equilibria $U^*$ of the classical model, which satisfy
(\ref{eq:symprop}). Using the fragile equilibria as a model of the low energy
classical states of the glass, where $\epsilon$ is identified with the thermal
equilibrium energy density, we expect $\epsilon\sim c_g/\beta$, where $c_g$ is
the low temperature limit of the classical specific heat. Finally, using
(\ref{eq:rhobeta}) we arrive at a Lifshitz-tail form for the transition state
density:
\begin{equation}
  \rho_0(\epsilon)\propto e^{-\epsilon_0/\epsilon}.
\end{equation}

We are now ready to consider the possibility of an Anderson transition. The
method we use is to estimate the fraction of the modes about a fragile
equilibrium, characterized by $\epsilon$, that are rubicons rather than
phonons. When a suitably large fraction are rubicons, the true quantum states
are delocalized. The critical $\epsilon$ above which this happens is analogous
to a mobility edge and depends on the mass parameter $\mu$ of the quantum
model. When interpreting the result we should remember that $\epsilon$ is only
the classical part of the excitation energy.

About each fragile equilibrium $U^*$ the potential energy has the form
\begin{equation}
  \Phi (X;U^*)=\Phi(U^*)+\frac{N}{2}\,\mathrm{Tr}(X^T K^* X)+\cdots,
\end{equation}
where
\begin{equation}
  K^*=\frac{1}{\sqrt{N}}\,{U^*}^T\mathrm{sgn}(U^*)
\end{equation}
is symmetric by (\ref{eq:symprop}). Let $V$ be the orthogonal transformation to normal modes $X'=V^T
X V$, where $K'=V^T K^* V$ is diagonal. This does not change the form of the
kinetic energy (\ref{eq:laplacian}). Also, since
\begin{equation}
  (K^*)^T K^*=\frac{1}{N}\mathrm{sgn}(U^*)^T\mathrm{sgn}(U^*)
\end{equation}
is approximately the identity matrix (because $\mathrm{sgn}(U^*)$ is nearly
Hadamard), so is the diagonalized matrix $K'$. Each normal mode coordinate $X'$
therefore has zero-phonon wave function
\begin{equation}
  \exp\left(-\sqrt{\mu/2} \,N\,X'^2\right)
\end{equation}
with amplitude of order
\begin{equation}\label{eq:rubiconamp}
  \sqrt{N}\;\delta X'\sim \mu^{-1/4}.
\end{equation}
Finally, since the measure (\ref{eq:measure}) is invariant with respect to the
normal mode transformation, so is the  density of transition state matrix
elements $\rho_0$. This gives
\begin{equation}\label{eq:f}
  f\propto \mu^{-1/4} e^{-\epsilon_0/\epsilon}
\end{equation}
as the fraction of the modes at a fragile equilibrium point that are rubicons,
where the proportionality factor is independent of $N$.

If the quantum-HMM has a mobility edge for $\epsilon>\epsilon_c$ we speculate
it occurs when the fraction $f$ exceeds a critical value, such that modes at
all the fragile equilibria are hybridized by rubicons to form extended
many-body states. Equation (\ref{eq:f}) then predicts $\epsilon_c$ vanishes
with the mass parameter only as
\begin{equation}
  \epsilon_c\propto\frac{1}{\log(1/\mu)},
\end{equation}
or that the low energy states do not thermalize when the quantum system is left
to itself. This is in contrast to the classical system coupled to a thermal
bath, where we saw no evidence of arrested dynamics. As in the classical case,
the low energy phenomenology places demands on the thermodynamic limit. In
particular, from (\ref{eq:rubiconamp}) we see that our mode analysis is well
defined only when $N$ becomes suitably large in the limit $\mu\to 0$ in order
to satisfy $\delta X'\ll 1$.

\section{Discussion}
\label{sec:conclusions}

The hard matrix model (HMM) challenges the hard sphere model (HSM) in its
privileged role of linking the worlds of physics and mathematics on the subject
of glass. Both models are highly symmetric and beset with ground states that
are largely mysterious, even after over a century of study. As a model of
glass, the HSM is considered in the limit of large dimensions
\cite{charbonneau2017glass} in addition to the usual thermodynamic limit of
many spheres. The HMM has only the matrix size $N$ in its thermodynamic limit.
Without quantum mechanics, by far the most studied case of hard spheres, each
model just has one parameter: the packing fraction for the HSM and the
temperature for the HMM. The molecular basis for both models --- packing of
particles (HSM) and covalent bond network (HMM) --- is tenuous because the
glass arises only in the infinite dimension limit.

Whereas both models are continuum models, the similarity ends already at the
level of the elementary modes. The HMM has true vibrations while the HSM
requires softening of the spheres (with additional parameters) to exhibit that
feature. An attractive feature of the HMM is the strong anharmonicity
introduced by the cusp in the local energy whenever one of the matrix elements
crosses zero. Phonons with significant content of such zero-crossing, called
rubicons, are believed to play a key role in the low energy dynamics. Their
density is proportional to the density of matrix elements near zero, $\rho_0$,
a structural property that was found to have a simple Arrhenius behavior.

The observation that $\rho_0$ is finite in the thermodynamic limit of the HMM
challenges the usual ``landscape" picture, where subsystems must receive energy
from a bath (the rest of the system) to surmount the energy barriers that carve
up the configurations into basins. Rubicons thus provide a mechanism whereby
the system can move between very different configurations (matrix elements
differing in sign) without activated processes. We believe rubicons are a
general feature of glassy systems and credit the HMM for highlighting their
existence.

The HMM offers many technical advantages over the HSM in simulations of
mechanics, thermodynamics, and even quantum mechanics. This is especially true
in the high dimension limit, where keeping track of imminent force
discontinuities (at zero-crossings) in the HMM equations of motion is easier
than forecasting the next collision event in the HSM. The relative ease of
thermodynamic MCMC simulations in the HMM allowed us to directly locate the
equilibrium phase transition to the Hadamard phase for sizes up to $N=20$. As
far as we know, liquid/crystal equilibrium has not been achieved with MCMC in
the HSM in dimensions above three. In the glass phase we found that $N$ had to
grow exponentially with the inverse temperature, in experiments up to $N=128$,
in order to see thermodynamic limit properties. We are not aware of HSM
simulations that go above nine dimensions.

The quantum extensions of the HSM and HMM are candidates for models without
disorder that exhibit many-body localization (MBL). It appears there is already
a 1D system that fits this description \cite{hickey2016signatures}. However,
there is no consensus on whether such systems might be generic, or whether MBL
is at all related to the phenomenology of classical glass. It would be daunting
to investigate the spectrum of the quantum-HSM (e.g. for evidence of a
level-statistics transition with packing fraction), and nothing close to such a
study has ever been attempted. On the other hand, localization for the
quantum-HMM Hamiltonian (\ref{eq:qmodel}) can easily and systematically be
investigated by tri-diagonalization \cite{haydock1981electronic} applied to a
Gaussian basis set with centers sampled near a fragile minimum.

\ack
JK-D is supported by NSF Grant No.~DMR-1719490. V.~E.\ thanks Persi Diaconis
for discussions. The authors thank James Sethna and David Huse for feedback on
the manuscript.

\appendix

\section*{Appendix}

\setcounter{section}{1}

This is a derivation of the equation of motion (\ref{eq:EOM}) for a particle moving on $\mathrm{SO}(N)$ in the presence of a potential $\Phi(U)$. We start with the Lagrangian
\begin{equation}
\mathcal{L}=\int dt\;\left(\frac{\mu N}{2}\,\mathrm{Tr}\,\dot{U}^T\dot{U}  -\Phi(U)+\mathrm{Tr}\,\Lambda^T(U^T U-1)\right),
\end{equation}
where all elements of $U$ are treated as independent variables while $\Lambda(t)$, a matrix of Lagrange multipliers, imposes orthogonality on $U$ at all times. Here are the Euler-Lagrange equations for $U$:
\begin{equation}\label{eom1}
\mu N \ddot{U}=-\nabla \Phi+U(\Lambda+\Lambda^T).
\end{equation}
The first term on the right hand side is just the matrix of partial derivatives of $\Phi$. Here is the transpose of the same set of equations:
\begin{equation}\label{eom2}
\mu N \ddot{U}^T=-(\nabla \Phi)^T+(\Lambda+\Lambda^T)U^T.
\end{equation}
Next, take one and two time derivatives of the constraint $1=U^T U$ to get two additional constraint equations:
\begin{eqnarray}
0&=&\dot{U}^T U+U^T\dot{U}\label{udot}\\
0&=&\ddot{U}^T U+2\,\dot{U}^T\dot{U}+U^T\ddot{U}\label{uddot}.
\end{eqnarray}
Now multiply (\ref{eom1}) on the left by $U^T$, (\ref{eom2}) on the right by $U$, add the result and use (\ref{uddot}) to eliminate the double time derivatives:
\begin{equation}
-\mu N\,\dot{U}^T\dot{U}=-\frac{1}{2}U^T(\nabla\Phi)-\frac{1}{2}(\nabla\Phi)^TU+\Lambda+\Lambda^T.
\end{equation}
Multiply this equation on the left by $U$ and take the resulting expression for $U(\Lambda+\Lambda^T)$ so obtained and substitute into (\ref{eom1}) to eliminate the Lagrange multipliers:
\begin{equation}
\mu N(\ddot{U}+U\dot{U}^T\dot{U})=-\frac{1}{2}\left(\nabla\Phi-U(\nabla\Phi)^TU\right).
\end{equation}
After using (\ref{udot}) to show $U\dot{U}^T\dot{U}=\dot{U}\dot{U}^TU$, we arrive at equation (\ref{eq:EOM}).

\section*{References}
\bibliographystyle{unsrt}
\bibliography{hmm}

\end{document}